\documentclass[10pt]{article}

\pdfoutput=1

\usepackage{geometry}
\usepackage{graphicx}
\usepackage{amsmath}
\usepackage{cite}
\usepackage{authblk}
\usepackage{hyperref}
\usepackage{xcolor}
\title{Explaining the geographic origins of seasonal influenza A (H3N2)}

\author[1]{Frank Wen\thanks{frankwen@uchicago.edu}}
\author[2]{Trevor Bedford} 
\author[1]{Sarah Cobey}
\date{\today}
\affil[1]{Department of Ecology and Evolution, University of Chicago, 1101 E. 57th St., Chicago, IL 60637}
\affil[2]{Vaccine and Infectious Disease Division, Fred Hutchinson Cancer Research Center}

\begin{document}

\maketitle

\section{Abstract}
Most antigenically novel and evolutionarily successful strains of seasonal influenza A (H3N2) originate in East, South, and Southeast Asia. 
To understand this pattern, we simulated the ecological and evolutionary dynamics of influenza in a host metapopulation representing the temperate north, tropics, and temperate south. 
Although seasonality and air traffic are frequently used to explain global migratory patterns of influenza, we find that other factors may have a comparable or greater impact. 
Notably, a region's basic reproductive number ($R_0$) strongly affects the antigenic evolution of its viral population and the probability that its strains will spread and fix globally: a 17-28\% higher $R_0$ in one region can explain the observed patterns. 
Seasonality, in contrast, increases the probability that a tropical (less seasonal) population will export evolutionarily successful strains but alone does not predict that these strains will be antigenically advanced. 
The relative sizes of different host populations, their birth and death rates, and the region in which H3N2 first appears affect influenza's phylogeography in different but relatively minor ways. 
These results suggest general principles that dictate the spatial dynamics of antigenically evolving pathogens and offer predictions for how changes in human ecology might affect influenza evolution.

\section{Introduction}
Antigenic variants of seasonal influenza continuously emerge and escape human immunity in a process known as antigenic drift. 
These drifted strains are less easily recognized by host immunity and therefore have a transmission advantage.
More antigenically advanced strains are also more likely to spread globally and successfully perpetuate the evolutionary lineage of subsequent variants.

Asia has long been recognized as a major source of not only new influenza subtypes but also new strains of seasonal influenza \cite{Shortridge:1982wo, Webster:1992wl, Cox:1994ui, Cox:2000gq}. 
Influenza A/H3N2, A/H1N1, and two B lineages currently circulate in the human population, with the H3N2 subtype causing the most disease \cite{FluNet}.
Phylogeographic analyses show that East, South, and Southeast Asia contribute disproportionately to the evolution of seasonal H3N2, exporting most of the evolutionarily successful strains that eventually spread globally \cite{Rambaut:2008ew,Russell:2008ke,Bahl:2011ep, Bedford:2010el, Bedford:2015fj}. 
The trunk of H3N2's phylogeny traces the evolutionary path of the most successful lineage and was estimated to be located in Asia 87\% of the time from 2000 to 2010 \cite{Bedford:2015fj}.
Additionally, strains of H3N2 isolated in E-SE Asia appear to be more antigenically advanced, with new antigenic variants emerging earlier in E-SE Asia than in the rest of the world \cite{Russell:2008ke, Bedford:2014bf}.
These observations suggest that ecological differences between regions, such as climate and human demography, affect the local antigenic evolution of H3N2, which in turn shapes its global migratory patterns.
Here we ask what ecological factors might cause disproportionate contributions of particular host populations to the evolution of an influenza-like pathogen.
This information may be immediately useful for viral forecasting. 
Over the long term, it could help predict changes in influenza's phylogeography and identify source populations to improve global vaccination strategies.

The conspicuous role of Asia in H3N2's evolution has been attributed to the seasonal nature of influenza in temperate regions \cite{Webster:1992wl, Viboud:2006ge, Rambaut:2008ew, Russell:2008ke, Bedford:2010el, Bahl:2011ep}.
Approximately 85\% of Asia's population and 48\% of the global population resides in a climatically tropical or subtropical region \cite{GPWv3} where semiconnected host populations support asynchronous epidemics that enable regional persistence year-round \cite{Viboud:2006ge, Russell:2008ke, Cheng:2013hb}. 
Uninterrupted transmission might increase both the efficiency of selection and the probability of strain survival and global spread.
By contrast, transmission bottlenecks from late spring through autumn in temperate populations necessarily limit local evolution and reduce opportunities for strain emigration \cite{Adams:2011dh, Zinder:2014jx}. 
Smaller contributions from other tropical and subtropical regions might arise from the weaker connectivity of their host populations \cite{Bedford:2010el, Chan:2010fu, Lemey:2014ez}.

Although seasonality clearly affects temporal patterns of viral migration \cite{Bahl:2011ep}, a robust explanation for differences in regions' long-term contributions to the evolution of H3N2 would consider the effects of seasonal variation in transmission in light of other potentially influential differences among host populations, including:

\textit{Host population size.}
E-S-SE Asia alone contains more than half of the global population \cite{UN:2013}. 
Larger host populations should sustain larger viral populations, and in the absence of other effects, they should contribute a proportionally larger fraction of strains that happen to spread globally.
Additionally, if rare mutations limit the generation of antigenic variants, larger populations could contribute a disproportionate number of antigenically novel strains with high fitness.

\textit{Host population turnover.} 
Birth rates have historically been higher in E-S-SE Asia than in most temperate populations \cite{UN:2013}. 
Demographic rates influence the replenishment of susceptibles and loss of immune individuals, thereby modulating selection for antigenic change. 
Faster replenishment of susceptibles increases prevalence, and thus viral abundance and diversity, but weakens the fitness advantage of antigenic variants.
A more immune population imposes greater selection for antigenic change but supports a smaller, less diverse viral population.  
Thus, the rate of antigenic evolution may vary in a complex way with the rate of host population turnover \cite{Grenfell:2004ho}. 

\textit{Initial conditions.}
H3N2 first emerged in or near Hong Kong in 1968.
The region in which a subtype emerges may effectively give the viral population a head start on evolution. 
The first epidemic will almost certainly occur in this region, and viruses here will be the first to experience selective pressure for antigenic change. 
If host migration rates are low and the founding viral population persists, this antigenic lead could be maintained or even grow in time. 

\textit{Transmission rates.}
Differences in human behaviour can affect transmission rates.
The transmission rate affects a strain's intrinsic reproductive number ($R_0$), the expected number of secondary cases caused by a single infection in an otherwise susceptible population.
Differences in regional $R_0$ could affect evolution in at least two ways.
Higher $R_0$ increases the equilibrium prevalence, increasing the probability that rare beneficial mutations will appear. 
In addition, the rate of antigenic drift increases with $R_0$ in models that include mutation as a diffusion-like process \cite{Lin:2003ht,Gog:2002bi, Kucharski:2012ez, Bedford:2015fj}. 
A higher intrinsic reproductive number in one population could thus accelerate the emergence of novel mutants in that area. 

To understand the potential effects of these five factors on the evolution of H3N2 in space, we simulated an influenza-like pathogen in a simplified representation of the global human metapopulation.
The simulated metapopulation consisted of three connected host populations, representing the temperate north, tropics, and temperate south.
Conceptually, the tropics in the model approximate Asia, where most of the population is tropical or subtropical \cite{GPWv3} and epidemics are asynchronous, and exclude other less connected tropical and subtropical populations on other continents\cite{Bedford:2010el, Chan:2010fu, Lemey:2014ez}. 
The two temperate populations approximate northern and southern populations where influenza is strongly seasonal. 
The model can also be generalized to represent three arbitrary populations by reducing seasonality.

We analysed the effects of these factors on two key metrics of influenza's spatial evolutionary and antigenic dynamics.
The first metric measures the proportion of the trunk of the phylogeny present in the tropics (figure~\ref{fig:example_run}\textit{a}).
The phylogenetic trunk represents the most evolutionarily successful lineage that goes on to seed all future outbreaks.
The second metric measures the degree to which tropical strains are antigenically advanced (figure~\ref{fig:example_run}\textit{b}).
Phenotypically, antigenic dissimilarities can be quantified as distances in antigenic space using pairwise measures of cross reactivity \cite{Smith:2004jc, Bedford:2014bf}.
Our model uses an analogous measure of antigenic distances, allowing us to determine the relative antigenic advancement of strains from each region.
We analysed these two metrics from simulations to test whether any of the five ecological factors could create spatial evolutionary patterns of a similar magnitude to the observed data.

\section{Results}
\subsection{Influenza-like patterns}
We simulated an individual-based model that included ecological and evolutionary dynamics in a metapopulation with three demes \cite{Bedford:2012bx}. 
By default, in one deme, transmission rates are constant throughout the year, and in the two others, transmission rates vary sinusoidally with opposing phases.
Viral phenotypes occur as points in 2D Euclidean space, and mutation displaces phenotypes in this 2D space according to a fixed kernel \cite{Bedford:2012bx}.
This space is analogous to an antigenic map constructed from pairwise measurements of cross-reactivity between influenza strains using a hemagglutination inhibition (HI) assay \cite{Smith:2004jc, Bedford:2014bf}.
Susceptibility to infection is proportional to the distance in antigenic space between the challenging strain and the nearest strain in the host's infection history, giving distant or antigenically advanced strains greater transmissive advantage.

\begin{figure}[h!]
\centerline{\includegraphics{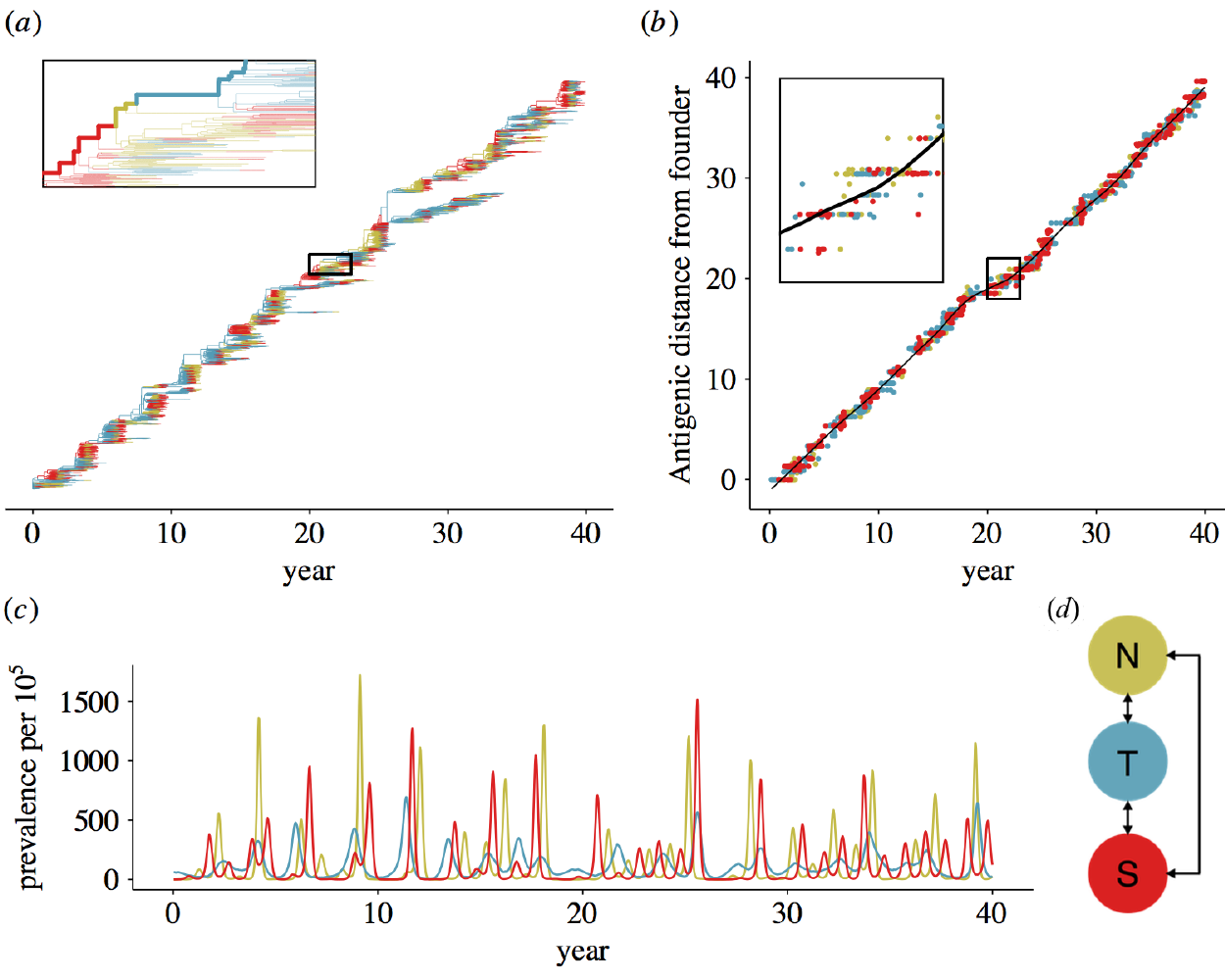}}
\caption{Representative output showing influenza-like behaviour from a sample simulation using the default parameters (table \ref{tab:default_parameters}). 
Statistics reported here are based on 53 replicate simulations. 
(\textit{a}) The phylogeny of the pathogen is reconstructed explicitly from the recorded ancestry of simulated strains. 
Branches are colored by region indicated in panel \textit{d}. 
The trunk is determined by tracing the recorded ancestry of surviving strains at the end of the simulation. 
Side branches show lineages that go extinct.
(\textit{b}) Viruses evolve antigenically away from the founding strain in a canalized fashion.
On average, the antigenic distance from the founding strain follows the trajectory indicated by the black LOESS spline fitted to viruses from all three regions.
At any given point in time, strains above this line have drifted farther from the founder compared to average, and are thus considered antigenically leading.
Conversely, strains below this line are considered antigenically lagging.
Antigenic lead is calculated as the distance to the spline in antigenic units.
(\textit{c}) Prevalence of infection over time for each region. 
(\textit{d}) Depiction of the totally connected model population, composed of the temperate north, tropics, and temperate south.} 
\label{fig:example_run}
\end{figure}

The model reproduces the characteristic ecological and evolutionary features of H3N2, except for the antigenic lead (table \ref{tab:flu_metrics}), under the default parameters (table \ref{tab:default_parameters}). 
We restricted our analyses to simulations where the virus remained endemic and where the time to the most recent common ancestor (TMRCA) never exceeded 10 years during the 40 years of simulation. 
We chose this cutoff because in some simulations, the viral population developed unrealistically deep branches.
In excluding extinctions and excessive diversity (branching), we assume that H3N2's historical evolutionary patterns represent the virus' likeliest evolutionary dynamics.
Of 100 replicate simulations, the viral population went extinct in 18 cases and exceeded the TMRCA threshold 29 times, leaving 53 simulations for analysis.
The model tracks the ancestry of individual strains, allowing us to explicitly reconstruct the phylogeny of the virus and the geographic location of lineages.
The phylogeny has the characteristically well-defined trunk with short branches of the H3N2 hemagglutinin (figure~\ref{fig:example_run}). 
This shape arises due repeated selective sweeps of antigenic variants, which reduces standing diversity; the average TMRCA across replicates was 3.72 years (SD = 0.26), comparable to empirical estimates of 3.89 years \cite{Bedford:2015fj}.
The antigenic distance from the founder increased linearly with time (figure~\ref{fig:example_run}), characteristic of H3N2's canalized antigenic evolution \cite{Bedford:2012bx,Smith:2004jc}. 
The mean antigenic drift across replicate simulations was 0.97 antigenic units per year (SD = 0.11), comparable to observed rates of 1.01 antigenic units per year \cite{Bedford:2014bf}. 
The mean annual incidence was 9.1\% (SD = 0.8\%). 
Reported annual incidence across all subtypes of seasonal influenza range from 9-15\% \cite{WHO:2014}.
Since we only modeled one lineage (e.g., the H3N2 subtype), the low estimate from the model is comparable to observed incidence.

Although all three host populations were the same size, the tropical strains were on average more evolutionarily successful. 
The phylogenetic trunk traces the most evolutionarily successful lineage and was located in the tropics 77\% (SD = 13\%) of the time, comparable to the observed 87\% of H3N2's trunk in E-S-SE Asia between 2000-2010 \cite{Bedford:2015fj}.
However, the default parametrization does not produce an antigenic lead in any population, despite the observed antigenic lead of Asian strains (table \ref{tab:flu_metrics}).
Antigenic cartography shows that while H3N2 drifts on average at 1.01 antigenic units per year globally \cite{Bedford:2014bf}, Asian strains tend to be farther drifted at any given time, and the region is thus considered to lead antigenically \cite{Russell:2008ke, Bedford:2014bf}.

\begin{table}[h!]
\centering
\caption{Properties of the default model}
\label{tab:flu_metrics}
\begin{tabular}{@{\vrule height 10.5pt depth4pt  width0pt}lrr}
\noalign{\vskip-11pt}\\
Statistic		&	Model mean $\pm$ SD	& 	Observed (Ref)\\
\hline
Annual incidence					&	0.091$\pm$0.0077	&	0.09 - 0.15 \cite{WHO:2014}\\
Antigenic drift rate (a.u. yr$^{-1}$)		&	0.97$\pm$0.11		&	1.01 \cite{Bedford:2014bf}\\
TMRCA (years)						&	3.7$\pm$0.26 		&	3.89 \cite{Bedford:2015fj}\\
Frac. of trunk in the tropics			&	0.61$\pm$0.13 		&  0.87 \cite{Bedford:2015fj}\\
Tropics antigenic lead (a.u.)			&	0.0025$\pm$0.036 	& 0.25 \cite{Russell:2008ke, Bedford:2014bf} \\
\end{tabular}
\end{table}

\begin{table}[h!]
\centering
\caption{Default parameters}
\label{tab:default_parameters}
\begin{tabular}{@{\vrule height 10.5pt depth4pt  width0pt}lrcccc}
\noalign{\vskip-11pt}\\
Parameter		&	Value	& Reference\\
\hline
Intrinsic reproductive number ($R_0$)	&	1.8		&	\cite{Jackson:2010hi, Biggerstaff:2014bu}\\
Duration of infection $\nu$				&	5 days	&	\cite{Carrat:2008bk}\\
Population size $N$					&	45 million	&	(see ESM) \\
Birth/death (turnover) rate $\gamma$	&	1/30 year$^{-1}$ & \cite{UN:2013} \\
Mutation rate $\mu$					&	$10^{-4}$ $\text{day}^{-1}$ & (see ESM) \\
Mean mutation step size $\delta_\text{mean}$	&	0.6 antigenic units & (see ESM) \\
SD mutation step size $\delta_\text{sd}$		&	0.3 antigenic units & (see ESM) \\
Infection risk conversion $c$			&	0.07		&	\cite{Bedford:2012bx, Gupta:2006bv, Park:2009jp}\\
Migration rate $m$					&	$10^{-3}$ $\text{day}^{-1}$ & (see ESM) \\
Seasonal amplitude $\epsilon$			&	0.10		&	\cite{Truscott:2012fh}
\end{tabular}
\end{table}

\subsection{Seasonality}
\begin{figure}[h!]
\centerline{\includegraphics{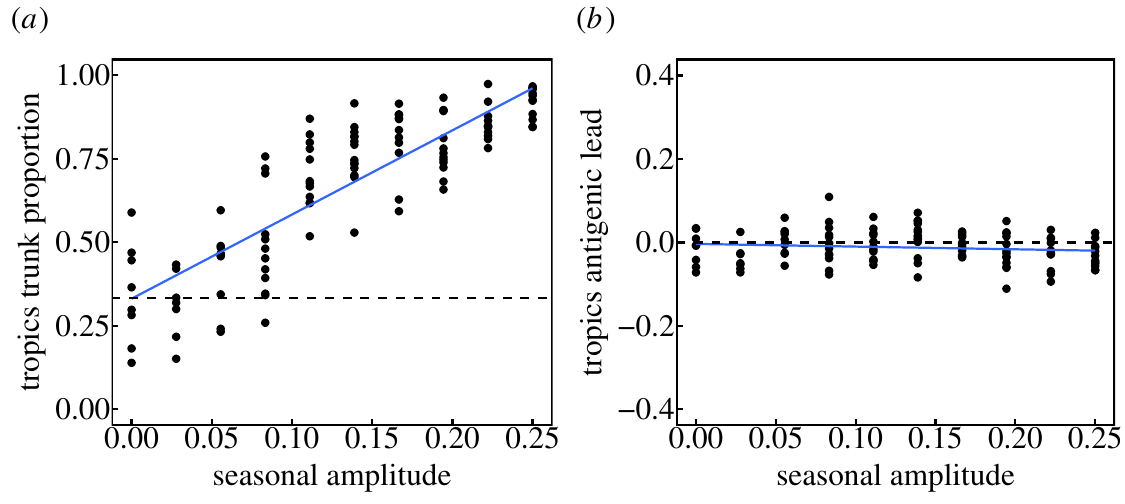}}
\caption{Seasonal amplitude $\epsilon$ in the temperate populations increases the tropics' contribution to the most evolutionarily successful lineage but alone does not affect regional differences in antigenic advancement. 
Transmission rates $\beta$ in the temperate north and south oscillate sinusoidally in opposite phase, with amplitude $\epsilon$. 
All other parameters remain at their default values (table \ref{tab:default_parameters}).
(\textit{a}) Effects of seasonality on the fraction of the trunk in the tropics (Pearson's $r = 0.85$, $p < 0.001$; $R^2 = 0.72$). 
Each point shows the fraction of time that the phylogenetic trunk was located in the tropics during the course of one simulation. 
The dashed line represents the null hypothesis where tropical strains comprise one third of the phylogenetic trunk. 
(\textit{b}) Effects on seasonality on the antigenic lead of the tropics (Pearson's $r = -0.12$, $p = 0.20$, $R^2 = 0.01$). 
Each point shows the average antigenic lead of tropical strains over time from one simulation. 
The dashed line represents the null hypothesis where tropical strains are neither antigenically ahead or behind. 
Blue lines represent linear least squares regression.}
\label{fig:seasonality}
\end{figure}

We first varied the strength of seasonal forcing, holding other parameters at their default values.
Seasonality by itself in the two temperate populations could not cause the tropics to produce more antigenically advanced strains; however, seasonality did cause the tropics to contribute a greater fraction of evolutionarily successful strains (figure \ref{fig:seasonality}). 
By linear regression, we estimate that the trunk would spend 87\% of its time in the tropics (the same fraction that is observed in Asia \cite{Bedford:2015fj}) with a seasonal transmission amplitude ($\epsilon$) of 0.19 (95\% CI: 0.18, 0.20).
Reduced seasonal forcing in the temperate populations equalized the fraction of the trunk in each population.
In multivariate sensitivity analysis, the amplitude of seasonal transmission accounted for 33\% of the variation in the tropical fraction of the trunk (electronic supplementary material, figure S2, table S2). 
This result suggests that seasonal bottlenecks in temperate populations discourage seasonal strains from fixing globally, in agreement with other models \cite{Adams:2011dh}. 
However, seasonality alone could not explain any variation in the tropic's antigenic lead (electronic supplementary material, figure S2, table S3).
We therefore hypothesized that ecological factors besides seasonality must contribute to regional differences in relative antigenic fitness.

\subsection{Transmission rate in the tropics}
\begin{figure}[h!]
\centerline{\includegraphics{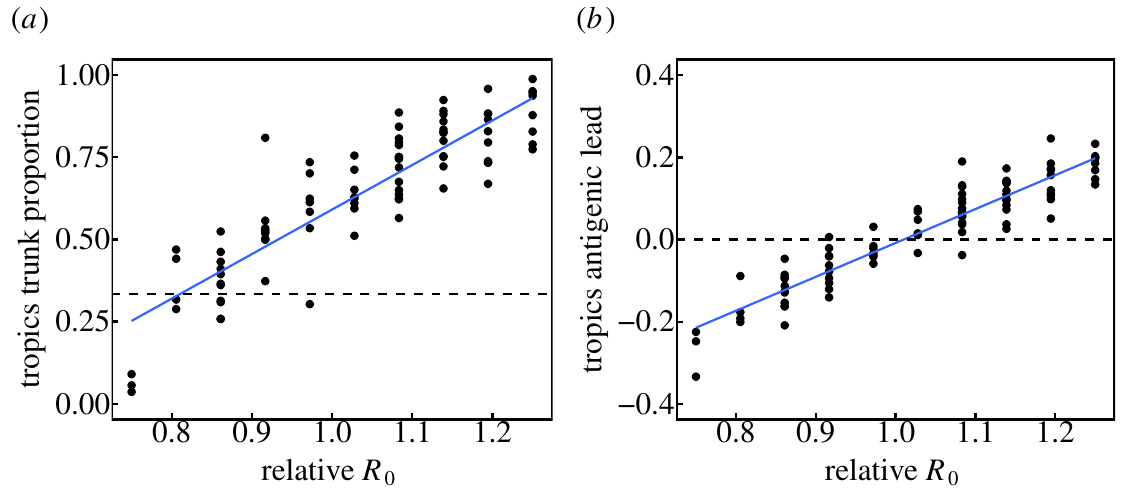}}
\caption{Increased $R_0$ in the tropics increases the tropics' contribution to the most evolutionarily successful lineage and the antigenic advancement of tropical strains. 
Relative $R_0$ is calculated as $R_{0}$ in the tropics divided by $R_{0}$ in the temperate regions. 
$R_0$ in the tropics was varied while $R_0$ in the temperate regions was kept at its default.
Other parameters were also kept at their default values (table \ref{tab:default_parameters}).
(\textit{a}) Effect of $R_0$ in the tropics on the fraction of the trunk in the tropics (Pearson's $r = 0.88$, $p < 0.001$; $R^2 = 0.78$). 
Each point shows the fraction of phylogenetic trunk located in the tropics during one simulation. 
The dashed line represents the null hypothesis where tropical strains comprise one third of the phylogenetic trunk. 
(\textit{b}) Effect of $R_0$ in the tropics on the antigenic lead in the tropics (Pearson's $r = 0.93$, $p < 0.001$; $R^2 = 0.87$).
Each point shows the average antigenic lead of tropical strains over time from one simulation.  
The dashed line represents the null hypothesis where tropical strains are neither antigenically ahead or behind. 
Blue lines represent linear least squares regression.}
\label{fig:relativeR0}
\end{figure}

Increasing $R_0$ in the tropics relative to the temperate populations caused the tropics to produce strains that led antigenically while also preserving the tropics' contribution to the trunk (figure \ref{fig:relativeR0}).
Linear regression implies that a 28\% (95\% CI: 25\%, 30\%) increase in $R_0$ in the tropics causes the tropics to produce strains that are, on average, 0.25 antigenic units ahead of global mean, reproducing the observed antigenic lead in Asia \cite{Russell:2008ke,Bedford:2014bf}. 
We also estimate that a 17\% increase in $R_0$ (95\% CI: 15\%,  19\%) causes the phylogenetic trunk to be located in the tropics 87\% of the time, reproducing the observed fraction of the H3N2 trunk in Asia \cite{Bedford:2015fj}.

The effects of $R_0$ on antigenic lead were robust to changes in other ecological variables and over a range of baseline values of global $R_0$.
When we varied the other parameters (table \ref{tab:default_parameters}), relative $R_0$ in the tropics accounted for 77\% of the variance in antigenic lead, making it the best predictor of antigenic lead in the tropics (electronic supplementary material, figure S2, table S3). 
The fraction of the trunk in the tropics also increased with the relative $R_0$, although $R_0$ explained less of the variation in trunk proportion (41\%), due to the effect of seasonality (electronic supplementary material, figure S2, table S2).

\begin{figure}[h!]
\centerline{\includegraphics{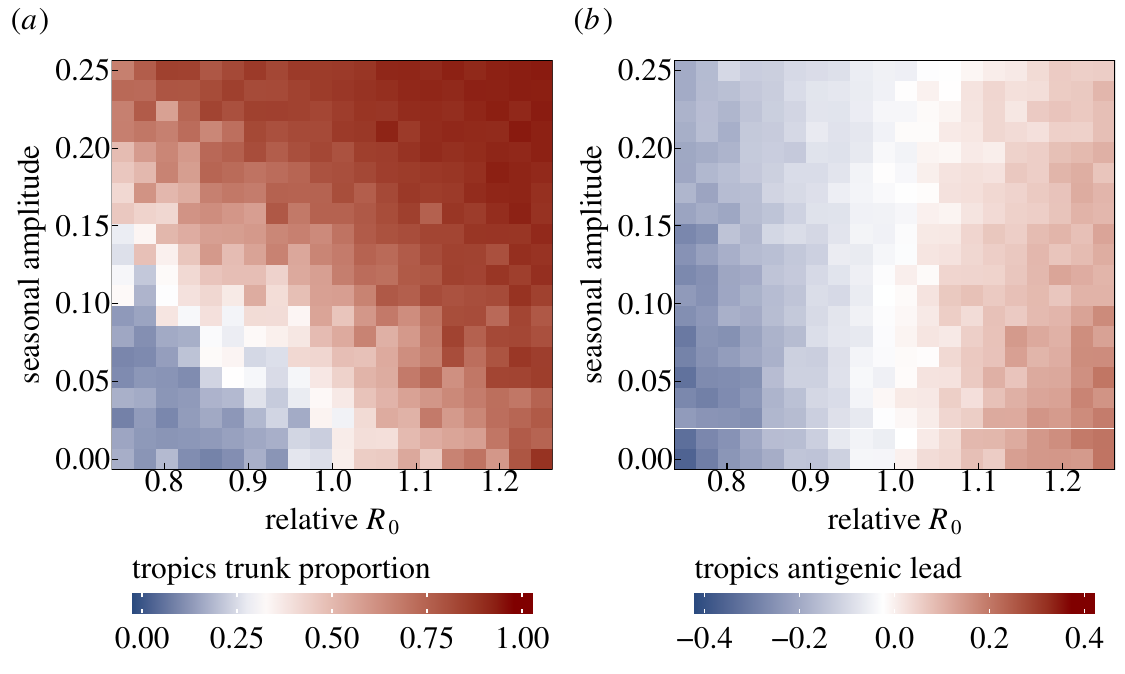}}
\caption{Seasonality in temperate populations has an equalizing effect on antigenic differences. 
Relative $R_0$ is calculated as $R_0$ in the tropics divided by $R_0$ in the temperate regions. 
(\textit{a}) Effects of seasonality and $R_0$ on the fraction of the trunk in the tropics. 
Blue indicates that the phylogenetic trunk is located in the tropics less than 1/3 of the time, and red indicates that the trunk is the tropics more than 1/3 of the time. 
(\textit{b}) Effects of seasonality and $R_0$ on antigenic lead in the tropics. Blue indicates that tropical strains are on average ahead antigenically relative to other global strains and red indicates that tropical strains are behind antigenically. 
Each square averages 1 to 17 replicate simulations.}
\label{fig:seasonality_relativeR0}
\end{figure}

Notably increased $R_0$ in one deme was sufficient by itself to make strains more evolutionarily successful and antigenically advanced. 
When we removed seasonality altogether to model three climatically identical populations, the population with the highest $R_0$ produced both the most antigenically leading and evolutionarily successful strains (figure \ref{fig:seasonality_relativeR0}).
Thus, higher $R_0$ alone in one region can cause it to attain an antigenic lead and fraction of the trunk as large as is observed in Asia.

To better understand why increasing regional $R_0$ causes that region to produce more antigenically advanced strains, we examined the effect of $R_0$ on antigenic evolution in a single deme.
Simulations showed that increasing $R_0$ increases the rate of antigenic drift (electronic supplementary material, figure S3).
To investigate further, we derived an analytic expression for the invasion fitness of a novel mutant in a population at the endemic equilibrium (electronic supplementary material, equation S1).
When the resident and mutant strains have the same intrinsic fitness ($R_0$), the growth rate of an antigenically distinct, invading mutant increases linearly with $R_0$ (electronic supplementary material, figure S4).
This linearity holds as long as the conversion between antigenic distance and host susceptibility (equation \ref{eq:risk_conversion}) is independent of $R_0$.
As $R_0$ increases, not only do mutants invade faster, but the invasion speed increases faster as a function of antigenic distance (electronic supplementary material, figure S4).

Although seasonality alone did not affect antigenic lead, the effects of $R_0$ on antigenic lead could be influenced by seasonality (figure \ref{fig:seasonality_relativeR0}).
Introducing seasonality in the temperate populations reduced differences in antigenic phenotype between regions. 
When tropical strains were antigenically ahead of temperate strains (due to higher tropical $R_0$), introducing seasonality reduced the tropics' antigenic lead. 
When tropical strains were antigenically behind temperate strains (due to lower tropical $R_0$), introducing seasonality reduced the antigenic lag.
Two factors explain the equalizing effect of seasonality on antigenic phenotype.
First, higher contact rates during transmission peaks in the two temperate populations increase the rate of strain immigration from the tropics.
Second, seasonal troughs in prevalence allow tropical strains to invade more easily due to reduced competition with local strains.

\subsection{Demographic rates, population size, and initial conditions}
Other ecological factors affected regional contributions to evolution but could not reproduce the observed patterns as well as differences in $R_0$ (electronic supplementary material, figures S1, S2). 
Notably, strains were slightly more antigenically advanced in older populations (electronic supplementary material, figure S1). 
When the rate of population turnover in the tropics was half that in the temperate regions, the tropics led by 0.04 antigenic units (SD = 0.03).
Larger populations generally contributed more to the trunk, although there was much variation that population size alone did not explain (electronic supplementary material, figures S1, S2 and table S2, S3). 
Initial conditions did not have a lasting effect (electronic supplementary material, figure S5).

\subsection{Implications for other influenza subtypes}
Influenza A/H1N1 and influenza B both evolve slowly compared to H3N2 and are suspected to have lower $R_0$ \cite{Bedford:2015fj, Bedford:2014bf}. 
Specifically, H1N1 drifts at a rate of 0.62 antigenic units per year, and the B/Victoria and Yamagata strains drift at 0.42 and 0.32 antigenic units per year respectively \cite{Bedford:2014bf}. 
H1N1 and B viruses are also less apt to have Asian origins than H3N2 \cite{Bedford:2015fj}. 
When we simulate with lower baseline $R_0$, we find that differences in $R_0$ between regions have a weaker influence on spatial patterns of evolution (electronic supplementary material, figure S8). 
Based on the relationship between mean $R_0$ and antigenic drift (electronic supplementary material, figure S3), we would expect seasonal H1N1, for example, to have an $R_0$ of 1.6. 
For this $R_0$, a 17\% increase in $R_0$ causes the tropics to occupy only 79\% (versus 87\% for H3N2-like $R_0$ of 1.8) of the trunk, and a 28\% increase in $R_0$ causes the tropics to lead by 0.20 (versus 0.25 for H3N2) antigenic units.

\section{Discussion} 

In our model, we find that the simplest explanation for why a host population produces more antigenically novel and evolutionarily successful strains than other populations is that its strains have a higher intrinsic fitness, or $R_0$. 
The strong effect of regional $R_0$ on spatial patterns of viral evolution is caused by the effect of $R_0$ on antigenic drift. 
Higher regional $R_0$ facilitates invasion of antigenically novel strains, resulting in faster antigenic drift.
Seasonality reduces the rate at which temperate populations export strains that are evolutionarily successful, but seasonality alone cannot explain regional differences in the production of strains that are antigenically novel. 
Size and age can influence global patterns too, but to a lesser extent: larger populations export more strains that fix, and populations with slower replenishment of susceptibles increase the rate of antigenic evolution. 
These last two effects are sensitive to changes in seasonality and $R_0$. 
These results highlight the relationship between human ecology and influenza's phylogeography.
Regions with high transmission rates may be expected to contribute disproportionately to influenza's evolution and may also be ideal targets for vaccine campaigns.
Accordingly, changes in human ecology can be expected to alter influenza's phylogeography.
These generalizations assume that H3N2 will evolve mostly as it has, with high strain turnover and limited genetic variation at any time, but more complex dynamics may be possible.

To make general predictions, we used a simple model. 
Although our three-deme metapopulation prevents us from replicating influenza's phylogeographic dynamics precisely, the model nonetheless reveals how ecological differences in populations create spatial patterns in the evolution of an influenza-like pathogen.
Simulations with more complex metapopulation models showed the same trends as the simple three-deme model (figures S9, S10), suggesting that our results are robust to changes in metapopulation population structure.

These results immediately raise the question of whether there is evidence of regional variation in $R_0$.
Low reporting rates and antigenic evolution make the $R_0$ of influenza difficult to measure with traditional methods, but we can conjecture from several lines of evidence.
Low absolute humidity favors transmission via aerosol in experimental settings \cite{Shaman:2009gp} and influences the timing of the influenza season in the United States \cite{Shaman:2010aa}. 
Based on absolute humidity and aerosol transmission alone, these results suggest that $R_0$ of tropical and subtropical Asia would be lower than in temperate latitudes. 
However, in Vietnam the onset of influenza-like illness is associated with periods of high humidity \cite{Thai:2015fp}. 
This observation suggests that humidity is not the dominant driver of influenza transmission, at least in this region.

Contact rates also influence transmission \cite{Wallinga:2006ct}.
Multiple studies have detected a significant effect of school closure on influenza spread \cite{Cauchemez:2008cq, Heymann:2004wy, Chao:2010fo}, although this trend is not without exception \cite{Cowling:2008bn}.
Households also influence risk: after one household member is infected, the average risk of secondary infection in a household contact is 10\% \cite{Tsang:2015kb}. 
Differences in classroom and household sizes may thus influence local transmission, and both are higher in, for instance, China and India than in Europe and the U.S. \cite{UNSD:2011, OECD:2015}.
Contact surveys report higher contact rates in Guangdong, China, than in European communities, whereas those in Vietnam are lower, although differences may arise from differences in survey design \cite{Read:2014kk, Mossong:2008kk, Horby:2011eh}. 
These surveys notably miss non-social, casual contacts (e.g., shared cafeterias and elevators) that might be important for influenza transmission.

Differences in local transmission rates may not scale: high rates of local transmission may be offset or attenuated by the structure of contact networks over larger areas. 
At the regional level, commuter and air passenger flows affect the spread of influenza epidemics, suggesting that adults are important to the long-range dispersal of the virus \cite{Viboud:2006ge, Lemey:2014ez}. 
The frequency of long-distance contacts differs between communities \cite{Read:2014kk}. 
Although sensitivity of $R_0$ to network topology is well known theoretically \cite{Miller:2009kq, Adler:1992wu}, there is a need to integrate the features of local and regional empirical transmission networks to infer large-scale differences $R_0$.  

Empirical estimates of $R_0$ are in theory attainable from seroprevalence.
Under a simplistic, single-strain $SIR$ model, which assumes random mixing and no maternal immunity, differences in $R_0$ should appear in differences in seropositivity by age.
For instance, if $R_0$=1.8, approximately 5.1\% of 2 year-olds would be seropositive, whereas 7.4\% would be seropositive if $R_0$ were 20\% higher.
$R_0$ variation in this range could be detected by sampling as few as 1500 2 year-olds in each population. 
Detailed surveys of H3N2 seropositivity by age cohort exist for some European countries \cite{Bodewes:2011fn, Sauerbrei:2009ic} but show much faster increases in seropositivity with age than expected under the $SIR$ model: 100\% of tested children are seropositive to H3N2 by age 7 in the Netherlands and by age 12 in Germany.
This discrepancy between theory and data may be due to antigenic drift resulting in higher attack rates \cite{Bedford:2015fj}.
The spatial difference in seroprevalence may also reflect greater contact rates among school-aged children \cite{Mossong:2008kk} and highlights the possibility that differences in exposure rates at young ages do not reflect mean differences in the populations. 
Such effects may be reduced by examining seroprevalence at older ages, but these estimates must balance a tradeoff between minimizing age-related correlations in transmission rates and increasing sample sizes required to detect asymptotically small differences in seropositivity. 
Another potential approach to measuring $R_0$ is to refine estimates of annual incidence in different populations.
Estimates of $R_0$ based on annual incidence would have to incorporate the histories of recent circulating strains, survey timing and titer dynamics, and vaccination in each population.

A greatly reduced birth rate confers a slight antigenic lead, but actual differences in birth rates between regions appear too small to explain Asia's observed lead. 
Current birth rates across most of Europe, China, and the United States are within 10\% of each other \cite{UN:2013}. 
Birth rates are almost twice as high in some SE Asian countries, including Cambodia, Laos, and the Philippines. 
The highest birth rates are found in Africa and the Middle East, and are three to four times higher than birth rates in the United States and China.
Our model suggests that these regions should contribute relatively less to influenza's antigenic evolution, assuming the differences in population structure are not associated with higher $R_0$, and ignoring other differences. 
However, taking age-assortative mixing into account may negate this expectation, with younger populations having increased $R_0$ \cite{Adler:1992wu, Dushoff:1995ht} thus contributing more to antigenic evolution.

We expect these results to apply to other antigenically varying, fast-evolving pathogens, including other types of influenza. 
Enterovirus-71 circulates globally, and its VP1 capsid protein experiences continuous lineage replacement through time, similarly to H3N2 hemagglutinin \cite{Tee:2010dia}.
Norovirus also demonstrates rapid antigenic evolution by amino acid replacements in its capsid protein \cite{Lindesmith:2008en}. 
We might expect that areas with high transmission contribute disproportionately to the antigenic evolution and global spread of these pathogens.
In addition, when we simulate with lower $R_0$, we find that differences in $R_0$ between regions influence spatial patterns of antigenic variation less (electronic supplementary material, figure S8).
This may explain why influenza A H1N1 and influenza B, which are suspected to have lower $R_0$ \cite{Bedford:2015fj, Bedford:2014bf}, are less apt to have Asian origins than H3N2 \cite{Bedford:2015fj}.

\section{Material and methods}

We implemented an individual-based $SIR$ compartmental model of an influenza-like pathogen, originally described by Bedford et al. \cite{Bedford:2012bx}. 
In this model, a global metapopulation is composed of three connected populations, representing tropics and temperate north and south. 
Individuals' compartments are updated using a $\tau$-leaping algorithm. 
Within a region $i$, the force of infection is given by 

\begin{equation}
F_{i}(t) = \beta_i (t) \frac{I_i}{N_i}
\end{equation}
where $I$ is the number of infected hosts. 
Between regions $i$ and $j$, the force of infection is given by

\begin{equation}
F_{ij}(t) = m\beta_j (t) \frac{I_i}{N_j} 
\end{equation}
where region $i$ is where the infection originates and region $j$ is the destination.
Here, $m$ is a scaling factor for interregional transmission, and $\beta_j$ is the transmission rate of the destination region. 
Transmission rates in the seasonal north and south oscillate sinusoidally in opposite phase with amplitude $\epsilon$. 
After recovery from infection, a host acquires complete immunity to viruses with that specific antigenic phenotype. 
Hosts that clear infection accumulate an infection history that defines their immunity. 
In a contact event, the distances between the infecting viral phenotype and each phenotype in the susceptible host's immune history are calculated. 
The probability of infection after contact is proportional to the distance $d$ to the closest phenotype in the host's immune history. 
An individual's risk of infection by such a strain is

\begin{equation}
\text{Risk} = \min\left\{1,cd\right\}
\label{eq:risk_conversion}
\end{equation}
where the proportionality constant for converting antigenic distance to a risk of infection $c=0.07$ \cite{Bedford:2012bx}; in other words, one unit of antigenic distance corresponds to 7\% reduction in immunity.
The linear relationship $c$ between antigenic distance and susceptibility derives from studies of vaccine efficacy \cite{Bedford:2012bx, Gupta:2006bv, Park:2009jp}.

Antigenic phenotypes are represented by points in a two-dimensional Euclidean antigenic space. 
One unit of antigenic distance in this space corresponds to a twofold dilution of antiserum in an HI assay \cite{Smith:2004jc}. 
The model is initialized at the endemic equilibrium with antigenically identical viruses. 
By default, all of the initial infections occur in the tropics.
Mutational events occur at a rate $\mu$ mutations per day. 
When a virus mutates, it moves in a random radial direction with a gamma-distributed step size. 
This mutation rate, along with the mutation size parameters ($\delta_\text{mean}, \delta_\text{sd}$) determine the accessibility of more distant mutations in antigenic space. 
The radial direction of mutation is chosen from a uniform distribution.

Additional methods are described in the electronic supplementary material. 

\section{Data accessibility}
Code implementing the model is available at \url{https://github.com/cobeylab/antigen-phylogeography.git}. 
The complete code for reproducing these results is available at \url{https://github.com/cobeylab/influenza_phylogeography_manuscript.git}.

\section{Competing interests}
We have no competing interests.

\section{Author contributions}
TB and SC conceived the study. FW performed the analysis and wrote the first draft of the paper. All of the authors contributed to and approve the final version.

\section{Acknowledgements}
We thank Daniel Zinder, Maciej Boni, and Greg Dwyer for helpful discussion and Ed Baskerville for programming guidance. 
This work was completed in part with resources provided by the University of Chicago Research Computing Center. 
SC was supported by NIH grant DP2AI117921.
FW was supported by NIH grant T32GM007281.

\makeatletter
\renewcommand*\thefigure{S\arabic{figure}}
\renewcommand*\thetable{S\arabic{table}}
\makeatother
\setcounter{figure}{0}
\setcounter{table}{0}

\section{Supporting Information}
\subsection{Extended methods}
\subsubsection{Selection of parameters}
Parameter values were selected to be consistent with influenza's biology and to reproduce its major epidemiological and evolutionary patterns (table 1).
The population size $N$ was chosen to minimize extinctions while also making efficient use of computational resources.
The population birth/death rate $\gamma$ = 1/30 year$^{-1}$ reflects the global crude birth rate estimates of 34 births per 1000 \cite{UN:2013}.

The proportionality constant $m$ for calculating the between-region contact rate was calculated from the number of international air travel passengers reported by the International Civil Aviation Organization divided by the global population \cite{ICAO:2014}.

We chose a baseline $R_0 = 1.8$. 
Estimates of $R_0$ from the first pandemic wave H3N2 in 1968 range from 1.06-2.06 \cite{Jackson:2010hi}, and estimates of $R_0$ for seasonal influenza range from 1.16 to 2.5, averaging approximately 1.8 \cite{Biggerstaff:2014bu}. 

The five-day duration of infection, $1/\nu$, is based on estimates from viral shedding \cite{Carrat:2008bk}. 
The transmission rate, $\beta$, is calculated using the definition of $R_0$:

\begin{equation*}
R_0 = \frac{\beta}{\nu + \gamma}
\end{equation*}
We chose the seasonal amplitude to ensure consistent troughs during the off-season in temperate populations while remaining within reasonable estimates of seasonal transmission rates \cite{Truscott:2012fh}.

Mutational parameters were selected to maximize the number of simulations where evolution was influenza-like (figure S6).
Mutations occur at a rate of $10^{-4}$ mutations per day.
This phenotypic mutation rate corresponds to 10 antigenic sites mutating at $10^{-5}$ mutations per day \cite{Rambaut:2008ew, Bedford:2012bx}.
The distance of each mutation is sampled from a gamma distribution with parameters chosen to yield a mean step size of 0.6 and a standard deviation of 0.3 antigenic units. 
These values correspond to a reduction in immunity of 4.2\% for an average mutation (SD = 2.1\%).
These mutation effect parameters give the gamma distribution an exponential-like shape, so that most mutations yield small differences in antigenic fitness, while occasionally mutations will yield greater differences. 
We chose $\mu$, $\delta_\text{mean}$, and $\delta_\text{sd}$ so that the simulations would exhibit influenza-like behaviour as consistently as possible (figure S6).
Here, the criteria for influenza-like behavior included endemism, reduced genealogical diversity (TMRCA $<$ 10 years) \cite{Bedford:2015fj}, and a biologically plausible mean rate of antigenic drift (1.01 antigenic units per year) \cite{Bedford:2014bf} and incidence (9-15\%) \cite{WHO:2014} (table 1, figure S6).

\subsubsection{Calculation of antigenic lead and trunk proportion}
We examined two metrics that describe influenza's evolutionary dynamics.
For computational tractability, these metrics were calculated using a subset of strains sampled over course of the simulation.
Strains were sampled proportionally to prevalence.

To calculate antigenic lead, we first calculated the antigenic distance of each sampled strain from the founding strain (figure 1\textit{a}).
We then fit a LOESS spline to these distances over time.
The spline describes the expected antigenic drift of circulating lineages at any point in time.
Strains above the spline have drifted farther than average and are considered antigenically leading.
Strains below the spline have drifted less than average and are considered antigenically lagging.
The antigenic lead in the tropics is calculated as the average antigenic distance to this spline for all sampled tropical strains.

To calculate the fraction of the trunk in each population, we first identified strains that comprise the trunk by tracing the lineage of strains that survive to the end of the simulation (figure 1\textit{b}). 
Because multiple lineages may coexist at the end of the simulation, we excluded the last five years of strains from trunk calculations.
The fraction of the trunk in the tropics is calculated as the fraction of the time the trunk was composed of tropical strains.

\subsubsection{Univariate sensitivity analysis}
In the univariate sensitivity analyses, we created regional differences in host ecology by varying each of the five ecological parameters individually ($R_0$, population turnover rate, seasonality, population size, and initial conditions) while keeping all other parameters at their default values (table 2).
To test the effects of regional $R_0$, we changed $R_0$ only in the tropics.
Similarly, we tested the effects of the rate of population turnover by varying it only in the tropics.
To investigate seasonality, we varied the seasonal amplitude of the transmission rate in the temperate populations.
(The transmission rate in the tropics was always constant over time.)
To explore population size, we examined the ratio of tropical to temperate population sizes, keeping the global population constant.
We initialized all simulations at the endemic equilibrium such that the total number of initial infecteds was constant.
We then scaled the number of initial infecteds in the tropics while keeping the number in the two temperate demes the same.
We ran twenty replicates for each unique combination of parameter values and discarded any simulations in which the virus went extinct or the TMRCA exceeded 10 years at any time in the 40-year simulation.
The analyses for antigenic lead and trunk proportion were performed on the remaining simulations (figure S1).

\subsubsection{Multivariate sensitivity analysis}
To test the robustness of the effects of individual parameters on the antigenic lead and the phylogenetic trunk, we simulated 500 points from a Latin hypercube with dimensions representing relative $R_0$, seasonality, relative population size, relative population turnover rate, and the fraction of initial infecteds in the tropics (figure S2). 
The ranges for each parameter (table S1) were chosen to remain within reasonable estimates. 
We simulated twenty replicates for each of the 500 unique parameter combination and discarded simulations in which the virus went extinct or the TMRCA exceeded 10 years at any time in the 40-year simulation.
We performed an ANOVA on the remaining 4119 influenza-like simulations to determine each parameter's contribution to the variance in antigenic lead and the fraction of the trunk in the tropics (table S2, S3).

\setlength{\parindent}{0cm}
\subsection{Invasion analysis}
We assume that the host population supports a resident strain at the endemic equilibrium. 
We develop an expression for the fitness of an invading mutant strain to explain how the selection coefficient of the mutant changes with $R_0$.\\\\
Here, $S,I$, and $R$ represent the fraction of susceptible, infected, and recovered individuals.
The birth rate $\gamma$ and the death rate are equal, so the population size is constant. 
All individuals are born into the susceptible class. 
Transmission occurs at rate $\beta$, and recovery occurs at rate $\nu$.

\begin{align*}
\frac{dS}{dt} &= \gamma (1-  S ) -\beta S I \\
\frac{dI}{dt}&= \beta S I - (\nu + \gamma) I\\
\frac{dR}{dt}&= \nu I - \gamma R
\end{align*}

We solve for the endemic equilibrium values of $S_\text{eq}$, $I_\text{eq}$, $R_\text{eq}$.

\begin{align*}
\frac{dI}{dt}=  0 &=\beta S_\text{eq} I_\text{eq} - (\nu + \gamma) I_\text{eq}\\
S_\text{eq} &= \frac{\nu + \gamma}{\beta} \equiv \frac{1}{R_0}
\end{align*}

$R_0$, the basic reproductive number, is defined as the number of secondary infections from a single infected individual in a totally susceptible population. 
Continuing to solve for $I_\text{eq}$ and $R_\text{eq}$, we have

\begin{align*}
\frac{dS}{dt}=  0 &=\gamma (1-  S_\text{eq} ) -\beta S_\text{eq} I_\text{eq} \\
I_\text{eq}&=\frac{\gamma}{\beta}(R_0-1)\\
\frac{dR}{dt} = 0 &= \nu I_\text{eq} - \gamma R_\text{eq}\\
R_\text{eq}&=\frac{\nu}{\beta}(R_0-1)
\end{align*}

To find the selection coefficient, we develop an expression for the effective reproductive number $R_e$ for both the resident and mutant strains. 
$R_e$ is the expected number of secondary infections from a single infected individual in a given population.
We will use the relationship

\begin{align*}
R_e &= SR_0
\end{align*}

The mutant strain is $d$ antigenic units from the resident strain. 
The conversion factor between antigenic units and infection risk is notated by $c$. 
Thus, the susceptibility to the mutant is given by $\min\{cd,1\}$, and immunity to the mutant is $\max\{1-cd,0\}$. 
For ease of notation, we assume $cd \le 1$, and use $k = 1-cd$.

The fraction of the population immune to the invading strain is denoted by $R'$. 
Note that the population is at the endemic equilibrium of the resident strain, and not the mutant.

\begin{align*}
R' &= (1-cd)R_\text{eq}\\
&= \frac{\nu}{\beta}(R_0-1)k
\end{align*}

We start by allowing coinfection. The fraction of susceptibles to the mutant strain is given by 

\begin{align*}
S' &= 1-R' - \frac{1}{N}\\
&= 1- \frac{\nu k }{\beta}(R_0-1) - \frac{1}{N}\\
\end{align*}

For large $N$, we have

\begin{align*}
S' &= 1- \frac{\nu k }{\beta}(R_0-1)\\
\end{align*}

As defined by our initial set of ODEs, the growth rates of the mutant and resident strains are

\begin{align*}
\frac{dI'}{dt} &= I' [\beta S' - (\nu + \gamma) ]\\
\frac{dI}{dt} &= I_\text{eq} [\beta S_\text{eq} - (\nu + \gamma) ]\\
\end{align*}

To get the selection coefficient, we take the difference between the growth rates:

\begin{align*}
s &= [\beta S' - (\nu + \gamma)] - [\beta S_\text{eq} - (\nu + \gamma) ]\\
&= \beta-\gamma k(R_0-1) - \frac{\beta}{R_0}
\end{align*}

Recall that $\beta = (\nu + \gamma) R_0$

\begin{align*}
s &= (\nu +\gamma)R_0 - \nu k (R_0-1) - (\nu+\gamma)\\
\end{align*}

Simplifying,

\begin{align}\tag{S1}
s&= (\nu cd + \gamma)(R_0 -1 ) 
\end{align}

Now disallowing coinfection, we have 

\begin{align*}
S' &= 1-R'-I_\text{eq}-I'\\
&= 1- \frac{\nu}{\beta}(R_0-1)k - \frac{\gamma}{\beta}(R_0-1) - \frac{1}{N}
\end{align*}

For large $N$,

\begin{align*}
S' &=1 - (R_0-1)[\frac{\nu k + \gamma}{\beta}] 
\end{align*}

Using the same arithmetic as in the case with coinfection, it follows that

\begin{align*}
s &= \beta-(\nu k + \gamma)(R_0-1) - \frac{\beta}{R_0}\\
\end{align*}

Simplifying,

\begin{align}\tag{S2}
s&=(\nu cd)(R_0-1)
\end{align}

In summary, the selection coefficient of an invading mutant strain increases linearly with the $R_0$, which is shared by both strains. 
The slope of this relationship is proportional to the distance $d$ between the two strains in antigenic space (figure \ref{fig:R0_selection}).
Naturally, relationship between the selection coefficient on the distance $d$ between strains depends on the functional relationship between antigenic distance and immunity.
However, the linear dependence of the selection coefficient on $R_0$ holds as long as the functional relationship between antigenic distance and immunity is independent of $R_0$. 

\subsection{Detecting differences in $R_0$}

In the $SIR$ model, the force of infection is

\begin{align*}
F = \beta I
\end{align*}

where $\beta$ is the transmission rate and $I$ is the fraction of infecteds. 
At the endemic equilibrium, the cumulative fraction of seropositive individuals at a given age $a$ is

\begin{align*}
f(a) &= 1-\exp(-\beta I_\text{eq} a)\\
&= 1-\exp(-R_0 (\nu + \gamma I_{\text{eq}} a)
\end{align*}

where $\nu$ is the recovery rate and $\gamma$ is the birth/death rate.
$I_{\text{eq}}$, the fraction of infecteds at the endemic equilibrium, is given by

\begin{align*}
 I_{\text{eq}} = \frac{\gamma}{\beta}(R_0-1)
\end{align*}

Figure \ref{fig:SIR_seropositivity} shows the fraction of seropositive individuals by age for the baseline $R_0=1.8$ and a 20\% higher $R_0=2.16$. 
The difference in the percentage of seropositive two-year-olds between the two groups is approximately 2.3\%.
The sample size in each group required to detect a difference $f_2(a) - f_1(a)$ with $\alpha$ confidence and $1-\beta$ power is

\begin{align*}
N =\frac{ f_1 (1-f_1) + f_2(1-f_2)}{(f_1 - f_2)^2} (\Phi_{\alpha/2} + \Phi_{\beta})^2
\end{align*}

For legibility, $f_i(a)$ is written as $f_i$. 
To detect a 20\% difference in $R_0$ between two populations with 0.05 significance and 0.80 power, we would require a sample of at least 1503 individuals in both groups.

\bibliography{agDriver}

\section{Supplemental tables and figures}
\begin{table*}[h!]
\centering
\caption{Parameter ranges used in Latin hypercube sampling}
\label{tab:LHS_parameters}
\begin{tabular}{@{\vrule height 10.5pt depth4pt width0pt}lrcccc}
\noalign{\vskip-11pt}\\
Parameter		&	Range\\
\hline
Relative $R_0$								&	0.8$-$1.2\\
Seasonal amplitude ($\epsilon$) in temperate populations 	&	0.0$-$0.15\\
Relative population size (N)					&	0.5$-$2.0\\
Relative turnover rate ($\gamma$)						&	0.5$-$2.0\\
Fraction of initial infecteds ($I_0$) in tropics		&	0.0$-$1.0\\
\hline
\end{tabular}
\end{table*}

\begin{table*}[h!]
\centering
\caption{ANOVA of the fraction of trunk in tropics from multivariate sensitivity analysis}
\label{tab:LHS_ANOVA_trunk}
\begin{tabular}{@{\vrule height 10.5pt depth4pt  width0pt}lrrrrrr}
  \hline
 Parameter & Df & Sum Sq & Frac of var & Mean Sq & F value & Pr($>$F) \\ 
  \hline
  Relative $N$         & 1 & 5.04 & 0.017 & 5.04 & 316.48 & $<$0.0001 \\ 
  Fraction $I_0$ in tropics  & 1 & 0.62 & 0.002 & 0.62 & 38.94 & $<$0.0001 \\ 
  Relative $R_0$        & 1 & 114.49 & 0.406& 114.49 & 7193.84 & $<$0.0001 \\ 
  Relative turnover  & 1 & 2.37 & 0.008 & 2.37 & 148.97 & $<$0.0001 \\ 
  Seasonal amplitude & 1 & 94.04 & 0.334 & 94.04 & 5908.50 & $<$0.0001 \\ 
  Residuals         & 4109 & 65.40& 0.232 & 0.02 &  &  \\ 
   Total         & 4114 & 281.96 & 1.000 & &  &  \\ 
   \hline
\end{tabular}
\end{table*}

\begin{table*}[h!]
\centering
\caption{ANOVA of the tropics' antigenic lead from multivariate sensitivity analysis}
\label{tab:LHS_ANOVA_agLead}
\begin{tabular}{@{\vrule height 10.5pt depth4pt width0pt}lrrrrrr}
\hline
 Parameter & Df & Sum Sq & Frac of var &Mean Sq & F value & Pr($>$F) \\ 
  \hline
  Relative $N$     & 1 & 1.94 & 0.033 & 1.94 & 754.71 & $<$0.0001 \\ 
  Fraction $I_0$ in tropics & 1 & 0.02 & $<$0.001 &0.02 & 8.75 & 0.0031 \\ 
  Relative $R_0$    & 1 & 44.55 & 0.766 & 44.55 & 17344.41 & $<$0.0001 \\ 
  Relative turnover & 1 & 1.04 & 0.018 & 1.04 & 406.81 & $<$0.0001 \\ 
  Seasonal amplitude & 1 & 0.02 & $<$0.001& 0.02 & 9.53 & 0.0020 \\ 
  Residuals & 4109 & 10.56 & 0.182& 0.00 &  &  \\ 
  Total         & 4114 & 58.140 & 1.000 & & &\\ 
   \hline
\end{tabular}
\end{table*}

\begin{figure*}[h!]
\centerline{\includegraphics{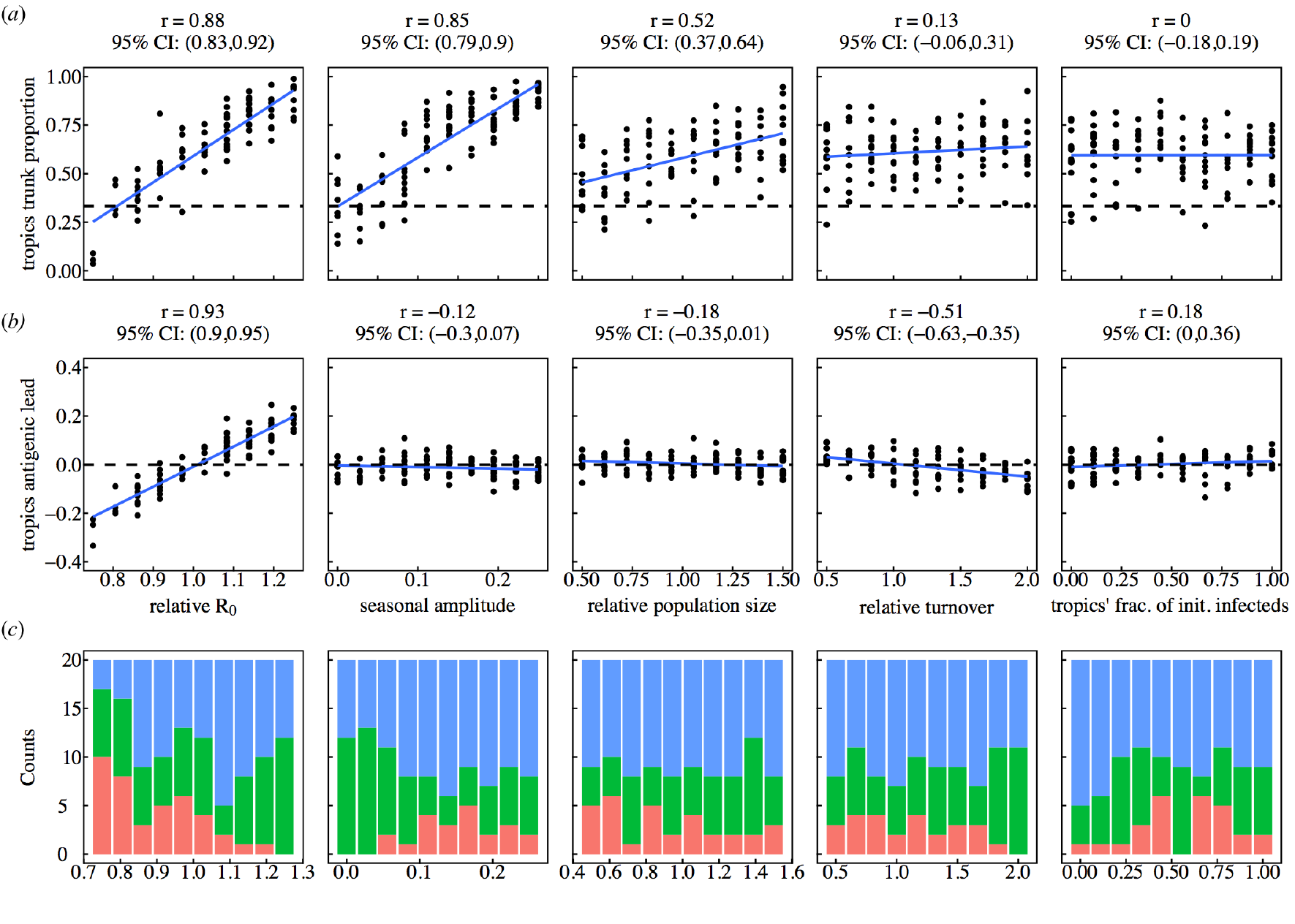}}
\caption{Univariate sensitivity analysis showing effects of individual parameters on (\textit{a}) the antigenic lead and (\textit{b}) the fraction of the phylogenetic trunk in the tropics. 
In each column of plots, only the parameter indicated on the x-axis is varying; all others are held constant at the default value. 
Each point represents the mean value over a single simulation. 
Blue lines indicate linear least squares regression. 
The dashed lines represent the null hypotheses where (\textit{a}) the trunk is distributed equally among the three regions or (\textit{b}) tropical strains are neither antigenically ahead or behind. 
(\textit{c}) Number of simulations that went extinct (red), exceeded the TMRCA limit (green), or were suitable for analysis (blue).}
\label{fig:univariate_summary}
\end{figure*}

\newpage
\begin{figure*}[h!]
\centerline{\includegraphics{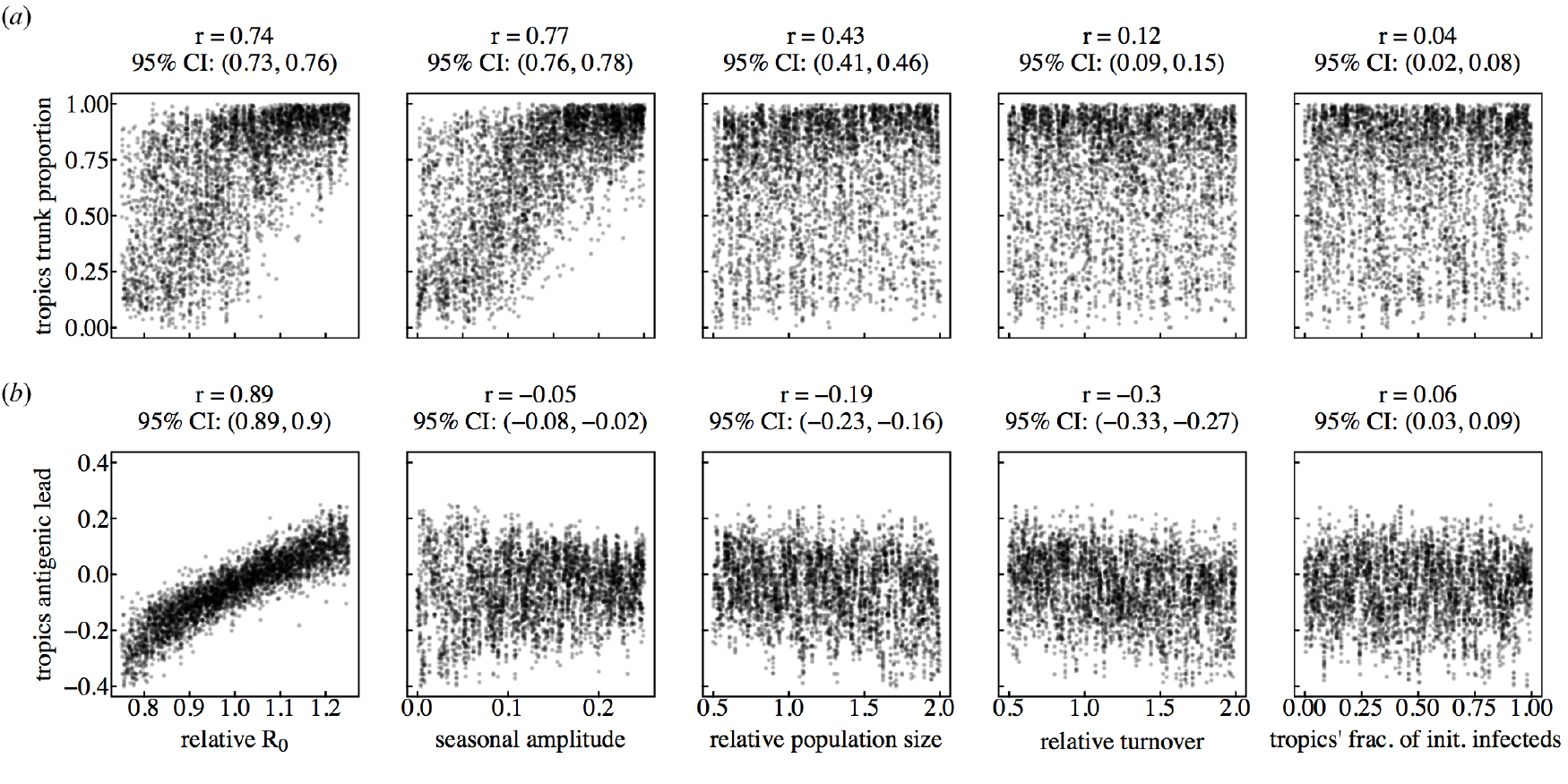}}
\caption{Multivariate sensitivity analysis showing effects of individual parameters on the (\textit{a}) antigenic lead and (\textit{b}) the fraction of the phylogenetic trunk in the tropics. 
Horizontal axes are projections of a Latin hypercube with dimensions corresponding to the five parameters indicated. 
Each point shows the mean value over a single simulation. 
The dashed lines represent the null hypotheses where (\textit{a}) the trunk is distributed equally among the three regions or (\textit{b}) tropical strains are neither antigenically ahead or behind. 
Pearson's correlation coefficients and associated 95\% confidence intervals are indicated.}
\label{fig:LHS_summary}
\end{figure*}

\newpage
\begin{figure*}[h!]
\centerline{\includegraphics{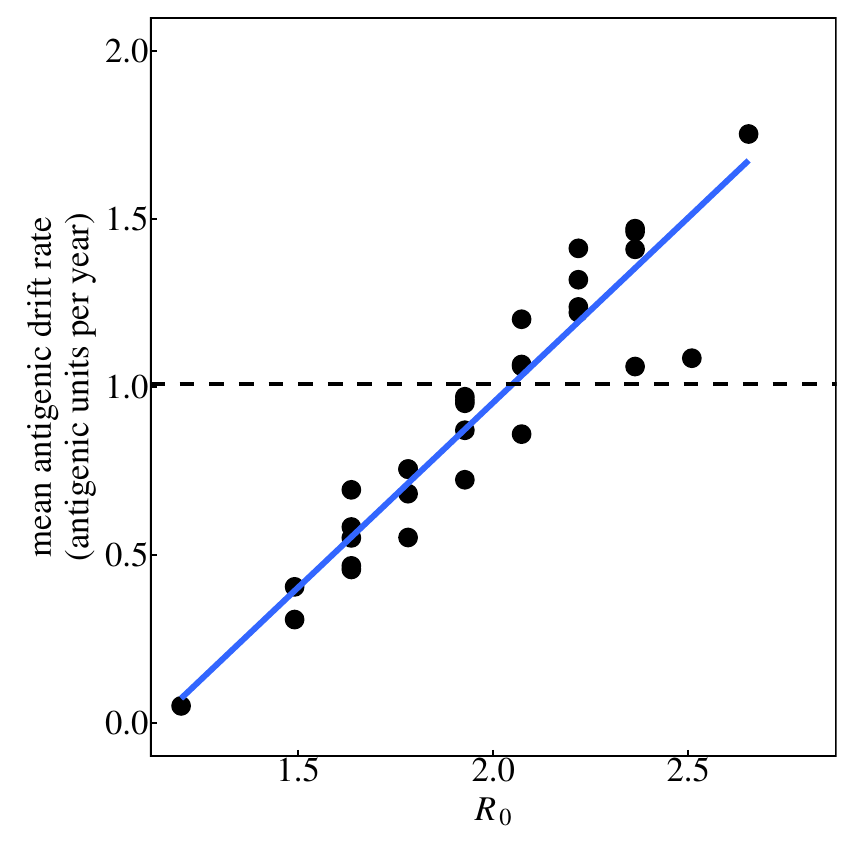}}
\caption{Effect of $R_0$ on antigenic drift in a single deme. 
Each point shows the mean antigenic drift rate from a single simulation. 
The blue line represents linear least squares regression, and the dashed line indicates the empirical estimate of the rate of antigenic drift for H3N2 \cite{Bedford:2014bf}. 
Pearson's $r = 0.94$, $p < 0.001$; 95\% CI: (0.88, 0.97).}
\label{fig:R0_drift}
\end{figure*}

\newpage
\begin{figure*}[h!]
\centerline{\includegraphics{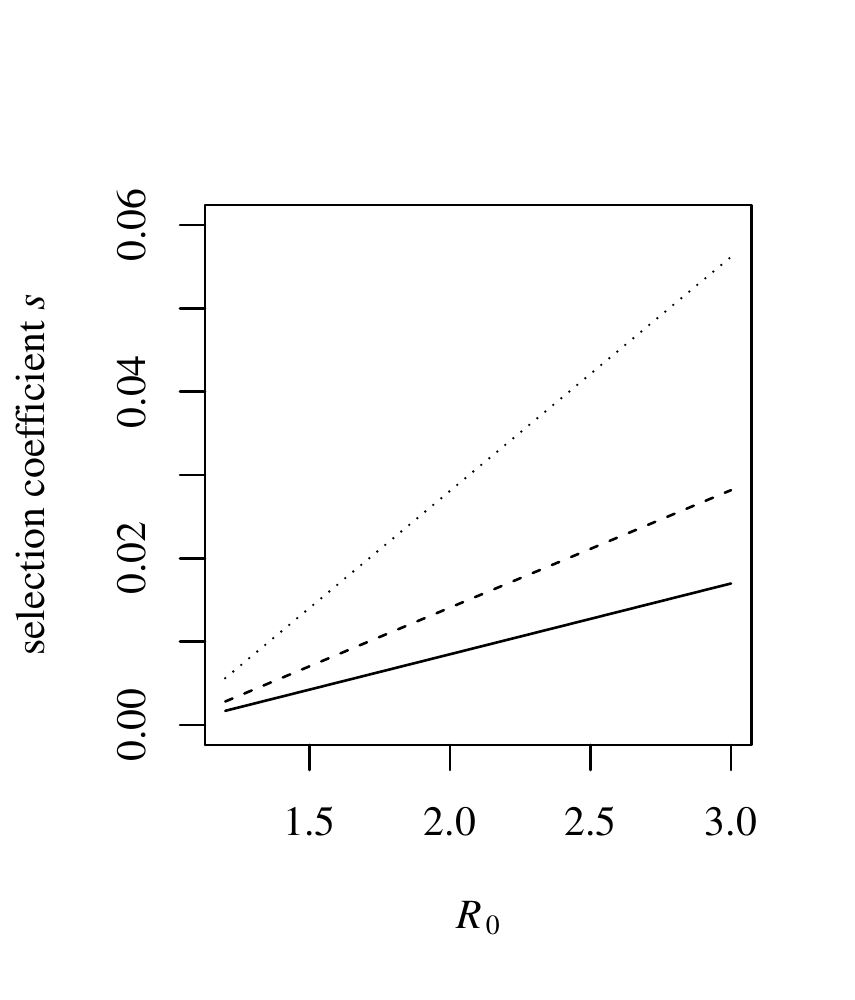}}
\caption{Relationship between $R_0$ and the selection coefficient for an invading strain with the resident at endemic equilibrium. 
The relationship for three different antigenic distances $d$ between the invading strain and the resident strain is shown: 0.6 (solid line), 1.0 (dashed line), or 2.0 (dotted line) antigenic units.}
\label{fig:R0_selection}
\end{figure*}

\newpage
\begin{figure*}[h!]
\centerline{\includegraphics{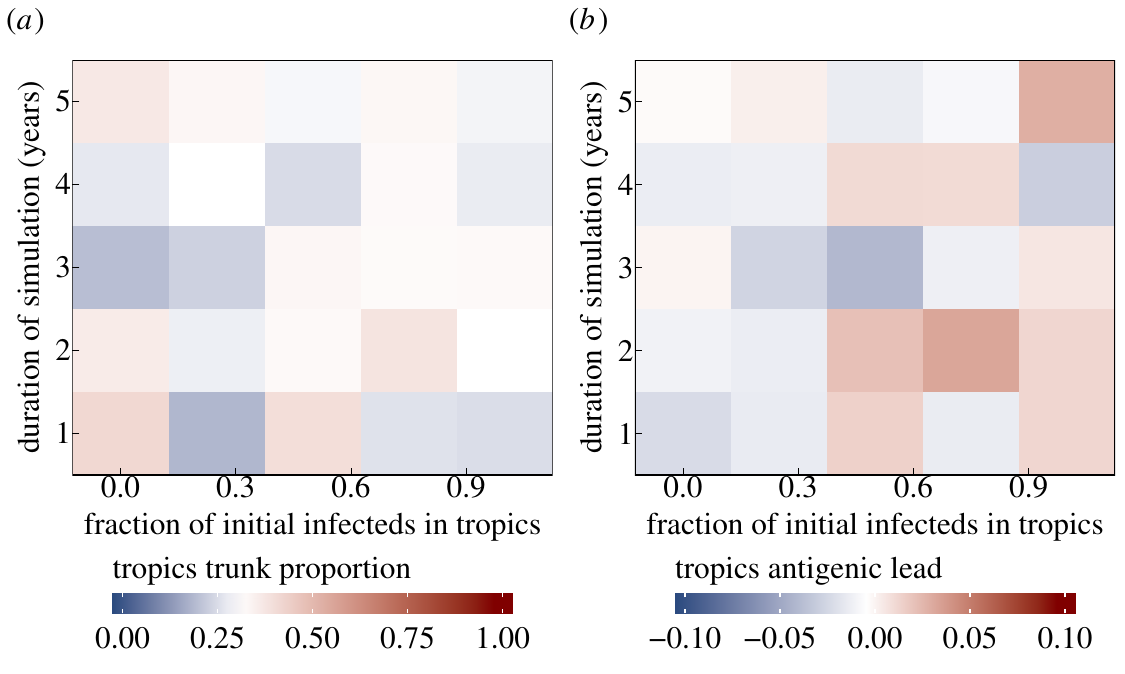}}
\caption{Effect of initial conditions on the antigenic lead and the fraction of the trunk in the tropics early in the simulation. 
Each square represents the average value for n=5 to 11 replicate simulations.}
\label{fig:I0_endYear}
\end{figure*}

\newpage
\begin{figure*}[h!]
\centerline{\includegraphics{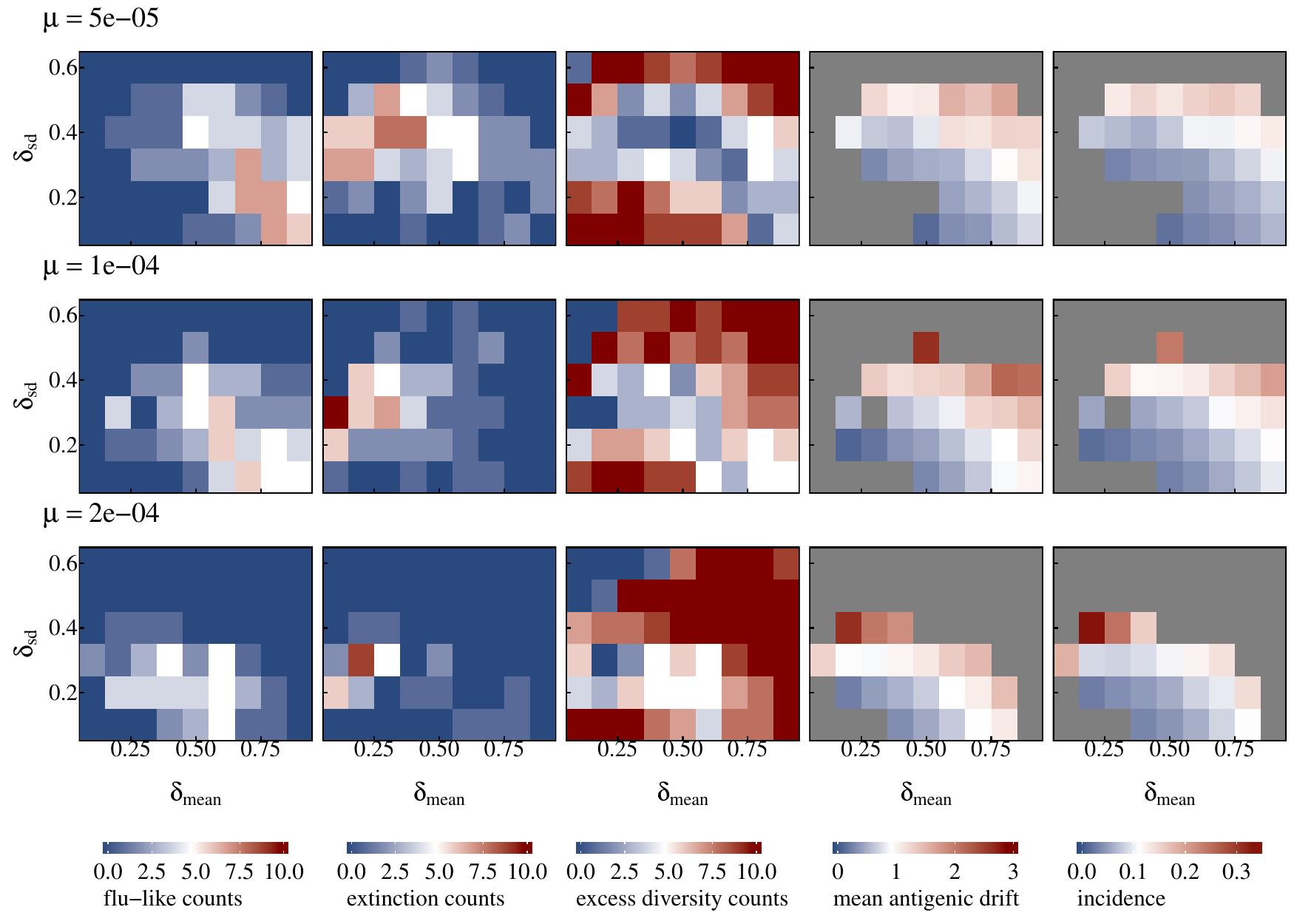}}
\caption{Sensitivity of influenza-like behaviour to changes in the mutational parameters, the mutation rate $\mu$, mean mutation size $\delta_\text{mean}$, and standard deviation of the mutation size $\delta_\text{sd}$. 
Within each plot, each square represents ten replicate simulations. 
Each row of plots shows results from simulations using different mutation rates $\mu$.
The number of simulations where the virus went extinct is shown in the second column of plots, and the number of simulations where the viral population exceeded a TMRCA of 10 years is shown in the third column of plots. 
The remaining simulations are considered influenza-like and are shown in the first column of plots. 
The reported mean antigenic drift rates and prevalences are averaged over the influenza-like replicates.
The color scales for mean antigenic drift and incidence are centered (white) at the  observed values for H3N2 (table 1).}
\label{fig:sensitivity_analysis}
\end{figure*}

\newpage
\begin{figure*}[h!]
\centerline{\includegraphics{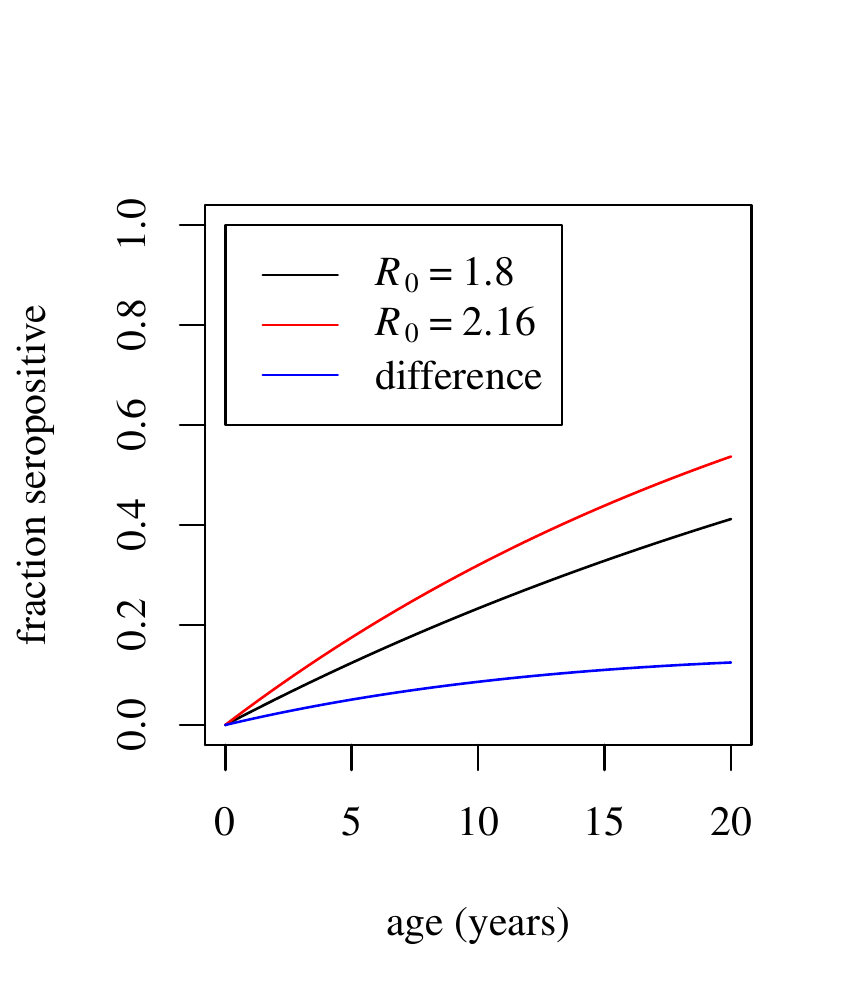}}
\caption{Theoretical increase in the fraction of seropositive individuals with age with $R_0=1.8$ and a 20\% higher $R_0 = 2.16$.}
\label{fig:SIR_seropositivity}
\end{figure*}

\newpage
\begin{figure*}[h!]
\centerline{\includegraphics{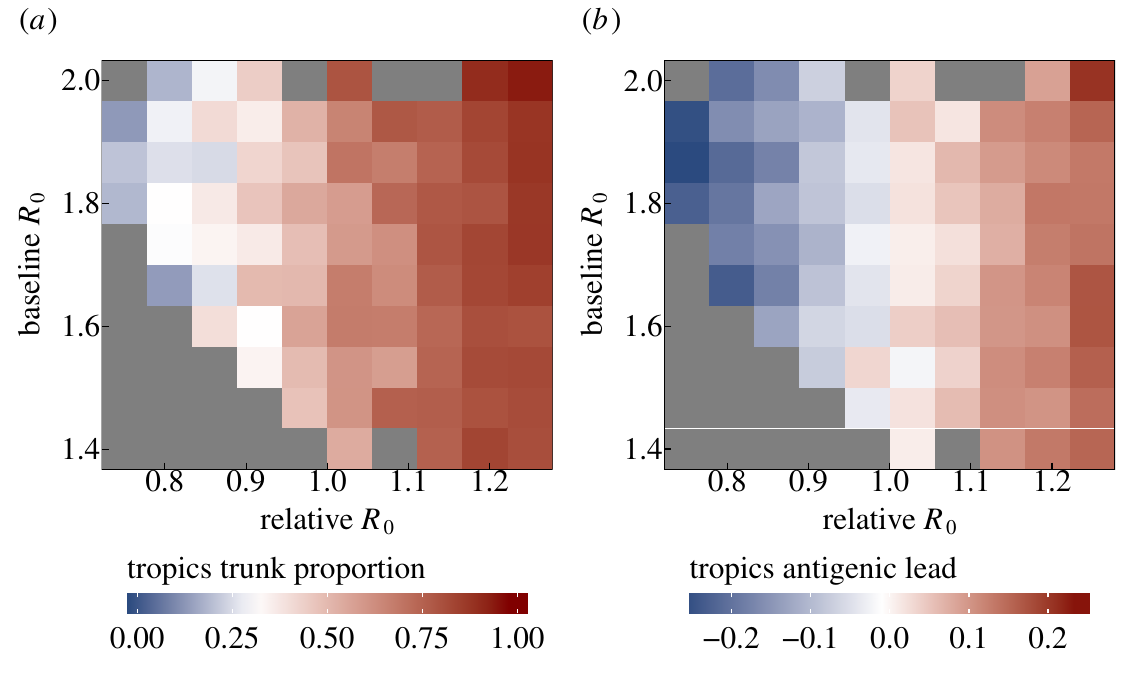}}
\caption{Lowering baseline $R_0$ decreases the effect of relative $R_0$ on the fraction of the trunk and antigenic lead in the tropics.
(\textit{a}) Effects of baseline and relative $R_0$ on the fraction of the trunk in the tropics. 
Blue indicates that the phylogenetic trunk is located in the tropics less than 1/3 of the time, and red indicates that the trunk is the tropics more than 1/3 of the time. 
(\textit{b}) Effects of baseline $R_0$ and relative $R_0$ on antigenic lead in the tropics. 
Blue indicates that tropical strains are on average ahead antigenically relative to other strains, and red indicates that tropical strains are behind antigenically. 
Each square represents an average from 1 to 14 replicate simulations.
Grey squares indicate parameter combinations where all of twenty attempted simulations either went extinct or exceeded the TMRCA threshold of 10 years.}
\label{fig:R0_relativeR0}
\end{figure*}

\begin{figure*}[h!]
\centerline{\includegraphics{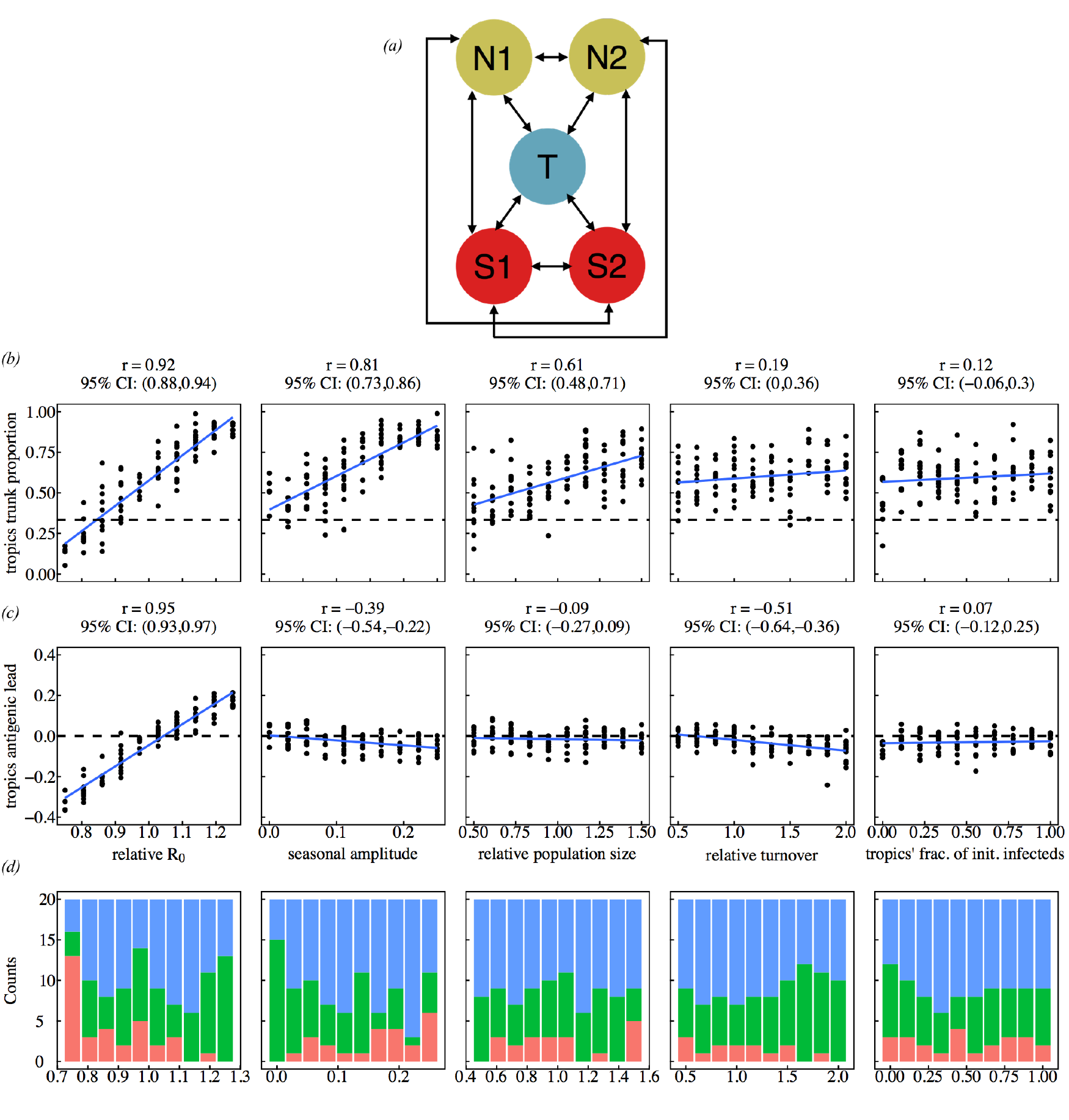}}
\caption{Univariate sensitivity analysis using a fully connected 5-deme model (\textit{a}) showing the effects of individual parameters on (\textit{b}) the antigenic lead and (\textit{c}) the fraction of the phylogenetic trunk in the tropics. 
By default, the tropics have a population size that is twice as large as any single temperate deme.
In each column of plots, only the parameter indicated on the x-axis is varying; all others are held constant at the default value. 
Each point represents the mean value over a single simulation. 
Blue lines indicate linear least squares regression. 
The dashed lines represent the null hypotheses where (\textit{b}) the trunk is distributed proportionally to the default population size among the regions or (\textit{c}) tropical strains are neither antigenically ahead or behind. 
(\textit{d}) Number of simulations that went extinct (red), exceeded the TMRCA limit (green), or were suitable for analysis (blue).}
\label{fig:univariate_summary_5_deme}
\end{figure*}

\begin{figure*}[h!]
\centerline{\includegraphics{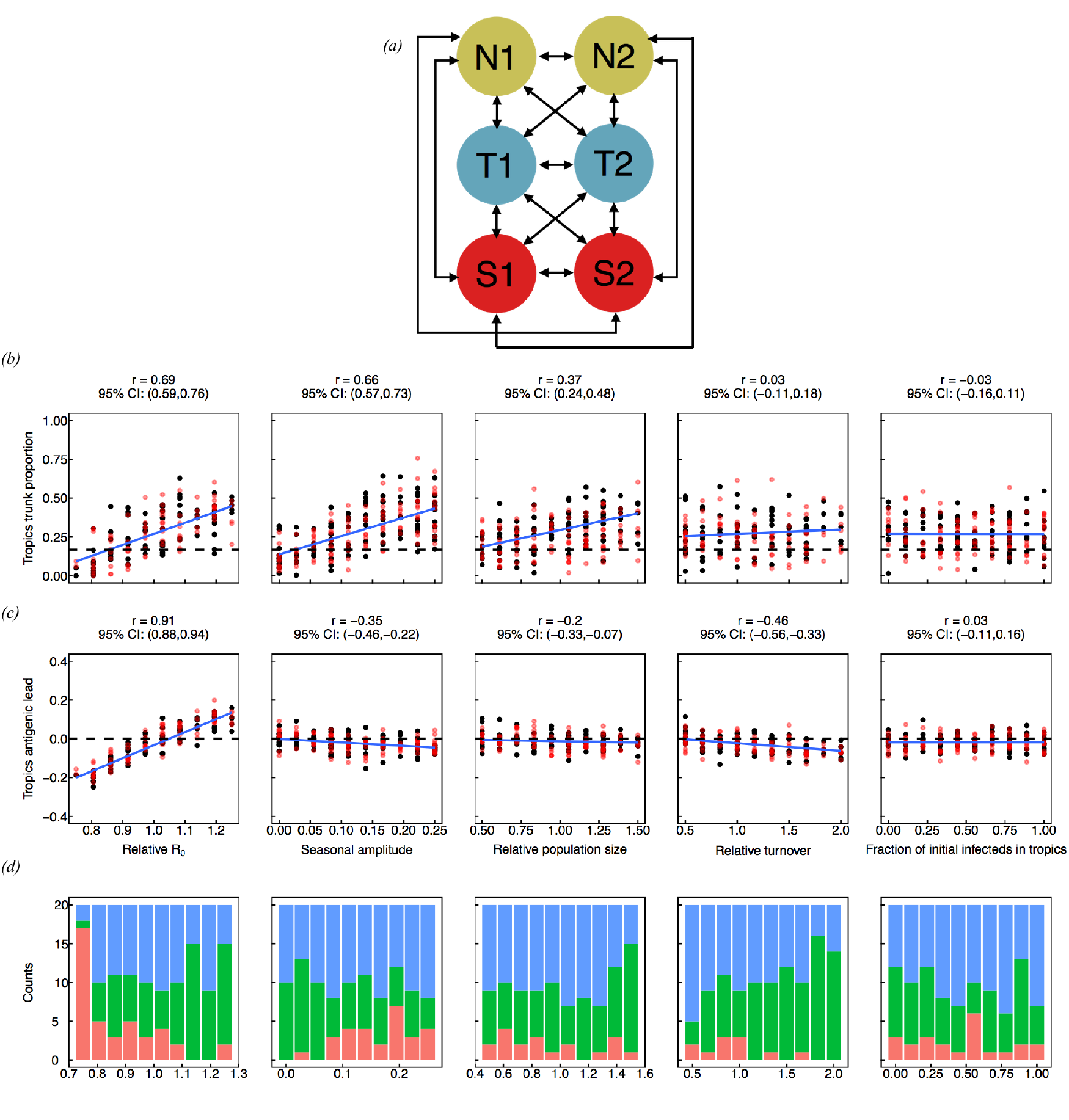}}
\caption{Univariate sensitivity analysis using a fully connected 6-deme model (\textit{a}) showing the effects of individual parameters on (\textit{b}) the antigenic lead and (\textit{c}) the fraction of the phylogenetic trunk in each of the two tropical demes. 
By default, all demes have the same population size.
In each column of plots, only the parameter indicated on the x-axis is varying; all others are held constant at the default value. 
Each point represents the mean value over a single simulation. 
Black points show results from one tropical deme and red points from the other.
Blue lines indicate linear least squares regression to the combined data from both tropical demes.
The dashed lines represent the null hypotheses where (\textit{b}) the trunk is distributed proportionally to the default population size among the regions or (\textit{c}) tropical strains are neither antigenically ahead or behind. 
(\textit{d}) Number of simulations that went extinct (red), exceeded the TMRCA limit (green), or were suitable for analysis (blue).}
\label{fig:univariate_summary_6_deme}
\end{figure*}

\end{document}